\documentstyle[prl,twocolumn,epsfig,aps]{revtex}
\begin{document}
\draft
\onecolumn 

\noindent

\title{ Directed Flow 
 in 158 A$\cdot$GeV $^{208}$Pb + $^{208}$Pb Collisions.}

\author{
M.M.~Aggarwal,$^{1}$
A.~Agnihotri,$^{2}$
Z.~Ahammed,$^{3}$
A.L.S.~Angelis,$^{4}$ 
V.~Antonenko,$^{5}$ 
V.~Arefiev,$^{6}$
V.~Astakhov,$^{6}$
V.~Avdeitchikov,$^{6}$
T.C.~Awes,$^{7}$
P.V.K.S.~Baba,$^{8}$
% ...
}

\author{(WA98 Collaboration)}

\address{$^{1}$~University of Panjab, Chandigarh 160014, India}
\address{$^{2}$~University of Rajasthan, Jaipur 302004, Rajasthan,
  India}
\address{$^{3}$~Variable Energy Cyclotron Centre,  Calcutta 700 064,
  India}
\address{$^{4}$~University of Geneva, CH-1211 Geneva 4,Switzerland}
\address{$^{5}$~RRC Kurchatov Institute, RU-123182 Moscow, Russia}
\address{$^{6}$~Joint Institute for Nuclear Research, RU-141980 Dubna,
  Russia}
\address{$^{7}$~Oak Ridge National Laboratory, Oak Ridge, Tennessee
  37831-6372, USA}
\address{$^{8}$~University of Jammu, Jammu 180001, India}
% ...
           
\date{\today}
\maketitle

\begin{abstract}
The directed flow of protons and $\pi^+$ have been
studied in 158 A GeV $^{208}$Pb + $^{208}$Pb collisions. 
A directed flow analysis of the rapidity dependence of the average
transverse momentum projected onto the reaction plane 
is presented for semi-central collisions with impact
parameters~$\approx 8$~fm, where
the flow effect is largest.  The magnitude
of the directed flow is found to be significantly smaller than observed at
AGS energies and than RQMD model predictions.

\end{abstract}

\twocolumn

Collective flow has been studied in heavy ion collisions since 
first observed at the Bevalac by the Plastic Ball experiment ~\cite{Gus84}.
At Bevalac energies of a few GeV per nucleon and lower, the study of collective
flow has been of interest largely due to its expected sensitivity to the
nuclear equation of state (EOS)~\cite{Sto86}. However, the extraction of
information on the EOS in heavy ion collisions is complicated 
by uncertainties in the initial dynamics of the pre-hydrodynamic stage, 
such as due to the momentum-dependence of the repulsive nucleon-nucleus 
interaction~\cite{Aic87} and possible in-medium modifications of the
nucleon-nucleon cross section~\cite{Rei97}.
Nevertheless, the importance of collective flow measurements 
in ultra-relativistic heavy ion collisions has been emphasized by
several authors~\cite{Ame91,Oll92,Hun95,Ris96,Sor97}.
Collective flow development follows the time evolution 
of pressure gradients in the hot, dense matter. Thus,
collective flow can serve as a hadronic ``penetrating probe''
to provide information on the initial state. In particular, the
formation of a Quark Gluon Plasma (QGP) during the early stages
of the collision is expected to result in reduced pressure gradients
due to a softening of the EOS, with a 
corresponding reduction of collective flow~\cite{Ame91,Hun95,Ris96}. 

Transverse collective flow
is normally discussed in terms of its lowest order symmetries with respect
to the reaction plane, which have recently been formulated in terms of
a Fourier decomposition~\cite{Vol96,Posk98}. 
The lowest order component is radial flow 
which is characterized by an isotropic transverse flow velocity.
The next component is the directed flow which is
characterized by the net displacement of the flow into a particular
transverse direction. The elliptic flow component corresponds to the 
second order Fourier coefficient of the flow pattern.

Recently, collective flow has been observed at the AGS energy of
11 A~GeV~\cite{Bar94}. The directed flow is observed to be 
smaller than at lower incident energy but similarly consistent with 
model calculations~\cite{Bar97}. It is also observed that
the elliptic flow has changed from an out-of-plane squeeze-out
direction to an in-plane direction~\cite{Bar97}.
At CERN SPS energies it has recently been shown that it is possible to 
determine the event-plane in the target fragmentation region for 
$^{16}$O- and $^{32}$S-induced reactions~~\cite{Awe96} 
and also in the mid-rapidity region
for $^{32}$S-~\cite{Agg97} and $^{208}$Pb-induced
reactions~\cite{Wie96}. 
Directed flow results at SPS energies were first reported
in~\cite{NisQGP} and both directed and elliptic flow results
have recently been published~\cite{App97}.
In this letter we
analyze the centrality dependence of the directed flow of protons
and $\pi^+$ in
158 A~GeV $^{208}$Pb + $^{208}$Pb and present a detailed analysis
of its rapidity dependence at the
intermediate centrality where it is greatest. 

The present analysis makes use of a subset of the detector systems of
the WA98 experiment. This subset consists
of the trigger detectors, the Plastic Ball detector,
and the tracking spectrometers. The
large aperture dipole magnet, Goliath, 
provided momentum analysis for the
tracking detectors. 
The data presented were  
taken during the 1996 SPS run period 
with 158 A GeV $^{208}$Pb beams using a 213~$\mu$m thick $^{208}$Pb 
target. The
WA98 minimum bias cross section for this run period, with magnetic
field on, was $\sigma_{mb} =$~6450 mb.

The trigger detectors consisted of
a nitrogen gas \v{C}erenkov counter to provide a 
fast beam trigger ($\leq~30$~ps time resolution), beam-halo veto counters, and the MIRAC calorimeter.
A beam trigger was defined as a signal in the start counter 
with no coincident signal in the veto counter
(which had a 3 mm diameter circular hole)
or in beam halo counters. 
The MIRAC measures the total transverse energy over the interval
$3.2 < \eta < 6.0$ with full azimuthal coverage over the
interval $3.7 < \eta < 4.9$. 
The WA98 minimum bias trigger requires a clean beam 
trigger with a MIRAC transverse energy which exceeds a low threshold.

The Plastic Ball detector provides full azimuthal coverage in the
target fragment region  (pseudorapidity $-1.7 < \eta < 0.5$) with 
$655$ detector modules.
It provides identification of pions, protons, deuterons, and tritons 
($\pi$, p, d, and t) with kinetic energies of 50 to 250 MeV by the 
${\rm \Delta E-E}$ method. In addition, stopped $\pi^+$
are identified by detection of the delayed $e^+$ 
from the decay $\pi^+ \rightarrow \mu^+ + \nu_{\mu} \rightarrow e^+ +
\nu_{e} + \overline{\nu}_{\mu}+ \nu_{\mu}$ in the Plastic Ball.
For the present analysis the rapidity region $-0.6 < y{\rm(proton)} < 0.3$
has been used.

The measurement of identified particles  near mid-rapidity was obtained using two 
large acceptance tracking arms beginning about 3.3 meters downstream
of the Goliath magnet. 
With the normal operation magnetic field setting the 
first tracking arm  was located to the side of
negative charge deflection and the second tracking arm to the
side of positive charge deflection. The positive charge results
are presented here. 
The  momentum resolution of the tracking system may be
parameterized as
$\Delta p/p \sim 0.97\%+0.16\%~p+0.023\%~p^2$ ($p$ in GeV/c). The acceptance for protons covered an interval 
$\Delta y_{p} \approx 0.3$ which shifted with transverse momentum
to provide  coverage over the region 
$ 1.4 < y_{p} < 2.4$. Particle identification
was obtained by time-of-flight measurement with 
a resolution of $< 90$ ps.

We determine the reaction plane as the azimuthal direction, $\Phi_0$,
opposite to $\vec{P_T}$, the total transverse momentum 
vector of fragments (p, d, and t) detected in the target rapidity 
region in the Plastic Ball detector.\footnote{Note: the opposite
direction is chosen by convention that the projectile fragments
define the direction of positive flow.}
To check for detector effects, mixed events are created
by mixing particles 
from different events keeping the same multiplicity distribution
as in the real events. The mixed events are then analyzed in the same
manner as the real events.
The laboratory distribution of 
$\Phi_0$ is nearly uniform,
with less than $2\%$ variations.  The $\Phi^{mix}$ distribution 
for mixed events is observed to show the same weak variations 
indicating small detector effects. The data have not been
corrected for this effect since the analysis of the mixed events shows
a negligible result.

\begin{figure}
 \begin{center}
    \epsfig{figure=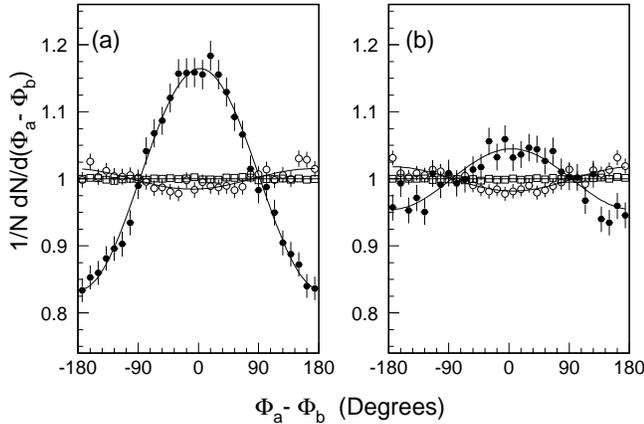,width=9.cm
 ,bbllx=0,bblly=0,bburx=550,bbury=320}
  \caption{The distribution of differences between
the total transverse momentum directions of two randomly chosen
equal size subevents of
fragments (p,d,t) in the target rapidity region for a) semi-central
(100 $< E_T <$ 200 GeV)  and b) central (380 $< E_T <$ 420 GeV) 
collisions of 158 A GeV $^{208}$Pb + $^{208}$Pb. 
Solid circles are for subevents within the 
same event. Open points are for subevents constructed from mixed
events. Solid curves are fits to guide the eye.
}
  \label{fig1}
 \end{center}
\end{figure}

In order to study how well the fragment flow direction is defined, 
we divide each event randomly into two equal sized subevents and determine
a fragment flow direction for each subevent,  $\Phi_a$ and $\Phi_b$.
If the direction of $\vec{P_T}$ is well-defined,
the directions determined from each subevent should be strongly 
correlated~\cite{Oll92,Dan85,Posk98}.
In Fig.~\ref{fig1}, the  $\Phi_a - \Phi_b$ correlation 
is shown for two different centrality bins.
As expected, the correlation observed for 
semi-central events is significantly larger than for very central 
events. Also shown are the results for mixed events allowing  (open squares)
or forbidding (open circles) multiple module hits. The former case 
demonstrates that detector non-uniformities are negligible.
In the latter case a
weak anti-correlation is observed due to the finite detector
granularity and an excluded-module effect. The data have been
corrected for this effect using such mixed events.

Azimuthal anisotropies of the particle emission are evaluated
by means of a Fourier expansion \cite{Vol96,Posk98}. 
The Fourier coefficients $v_n (n = 1,2)$ 
are extracted from the azimuthal distribution of identified particles 
with respect to the reaction plane, $\Phi_0$, which is determined
using all other fragments in the Plastic Ball.
\begin{eqnarray}
\frac{1}{N} \frac{dN}{d(\phi-\Phi_0)} 
= 1 + 2v'_1{\rm cos}(\phi-\Phi_0) + 2v'_2{\rm cos}(2(\phi-\Phi_0)),
\end{eqnarray}
where $\phi$ is the measured azimuthal angle.
The Fourier coefficient $v'_1$ quantifies 
the directed flow, whereas $v'_2$ quantifies the elliptic flow.
The coefficients must be corrected for the event plane resolution
as $v_n = v'_n / \langle {\rm  cos}(n(\Phi_0 - \Phi_r))\rangle$ where 
$\Phi_0 - \Phi_r$ is the deviation of the measured reaction plane from
the true reaction plane.  The event plane resolution may
be extracted from the correlation between subevents. For weak
correlations one expects
$\langle {\rm  cos}(\Phi_0 - \Phi_r)\rangle \simeq \sqrt{ 2 \langle {\rm
    cos}(\Phi_a - \Phi_b)\rangle}$. Using the more accurate
procedure and interpolation formula of Ref.~\cite{Posk98} one obtains
${\langle {\rm cos}(\Phi_0 - \Phi_r)\rangle} = 0.377\pm0.018$ for 
the semi-central (100 $< E_T <$ 200 GeV) event selection.

\begin{figure}
 \begin{center}
    \epsfig{figure=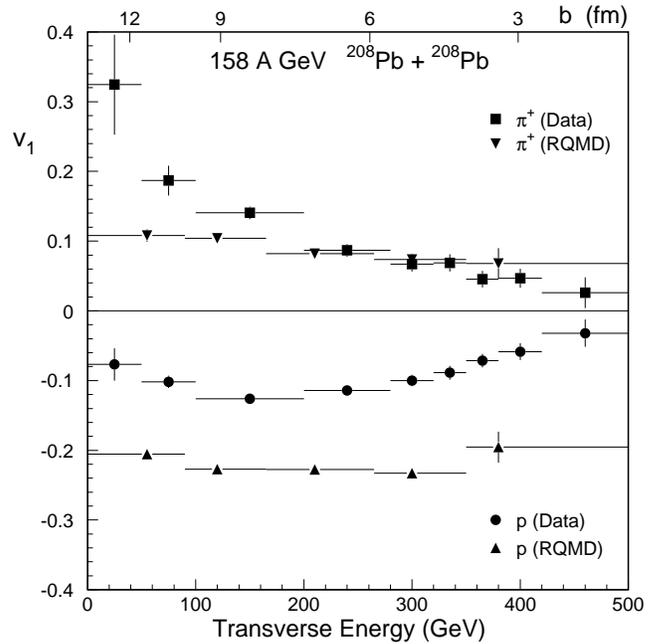,width=9.cm
 ,bbllx=0,bblly=0,bburx=550,bbury=500}
  \caption{The centrality dependence of the 
    directed flow coefficient $v_1$ for protons (circles) and $\pi^+$
    (squares). Triangles are results from RQMD  model calculations.
    The data have been corrected for the event-plane
    resolution. The vertical bars indicate the uncertainty of the
    fit and resolution correction.
    The horizontal bars indicate the $E_T$ bin intervals
    (or impact parameter intervals for RQMD).}
  \label{fig2}
 \end{center}
\end{figure}

The dependence of the $v_1$ fit parameter on
centrality, as determined by the measured transverse energy $(E_T)$,
is shown in Fig.~\ref{fig2}. 
For convenience an impact parameter scale is also shown. The $E_T$
scale has been converted to an impact parameter scale assuming a
monotonic relationship between the two quantities, and equating
$d\sigma/dE_T$ with $d\sigma/db$. 
As seen in Fig.~\ref{fig2},   the strength of the
directed flow of protons increases with
centrality and reaches a maximum value for semi-central collisions 
with $b \approx 8$ fm.  
It is interesting to note that the strongest flow effect
occurs at larger impact parameters than observed at lower
incident energy for similar systems 
(where $b \approx 4$ fm)~\cite{Rei97,Bar94}. 
For comparison, RQMD~2.3~\cite{Sor90} model predictions 
are shown subjected to the
same analysis after applying the Plastic Ball detector acceptance,
but using the true reaction plane.
RQMD predicts a significantly stronger correlation for protons than 
observed.

Also shown in Fig.~\ref{fig2} is the strength of the directed flow
of $\pi^+$,  identified in the Plastic Ball. A clear anti-correlation,
or anti-flow~\cite{Jan94}, is observed between the fragment and  $\pi^+$ flow
directions.  This behaviour has been observed at incident energies
from 1~A~GeV to SPS energies and has been 
explained as resulting from preferential absorption of the pions
emitted in the target spectator direction~\cite{Bar97,Awe96,Jan94,Kug94}. 
The absorption results in an oppositely directed apparent $\pi^+$ flow. 
The strength of the anti-correlation increases for the most 
peripheral events, indicating the increasing role of absorption.

A conventional directed flow analysis has been performed~\cite{Dan85},
in which the average transverse momentum
with respect to the reaction plane $\langle p_x\rangle$ is evaluated as a function
of rapidity. 
This is done for  semi-central collisions (100 $< E_T <$ 200 GeV)
where the largest azimuthal asymmetry is observed (see
Fig.~\ref{fig2}).  
The distribution $d^3N/dp^{\prime}_xdp^{\prime}_ydy$ is constructed
for protons and $\pi^+$ in the Plastic Ball and in the tracking arm, where 
the new axis $p^{\prime}_x$ corresponds
to the reaction plane determined event-by-event using all remaining 
fragments measured in the Plastic Ball (then reflected, $p^{\prime}_x
\rightarrow -p^{\prime}_x$, to correspond to the projectile fragment
direction, according to convention).
At each rapidity the average transverse momentum in the reaction plane,
$\langle p^{\prime}_x\rangle$, is calculated from fits to the experimental
distributions. 

\begin{figure}
 \begin{center}
    \epsfig{figure=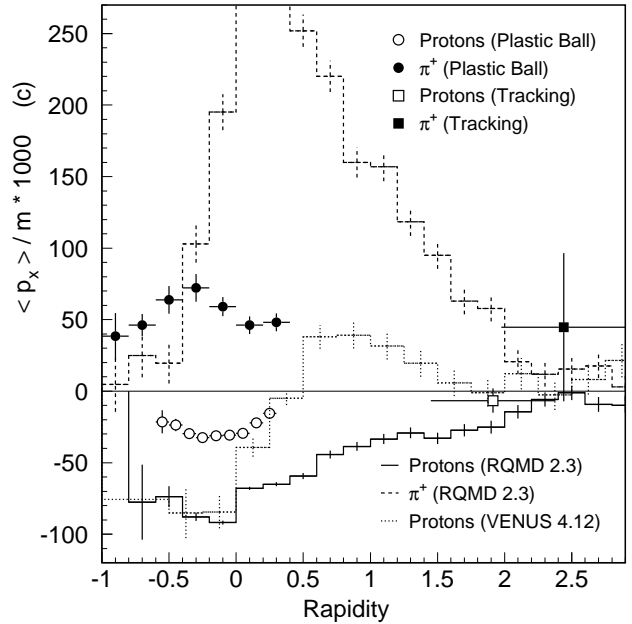,width=9.cm
 ,bbllx=0,bblly=0,bburx=550,bbury=500}
  \caption{ The average transverse momentum 
    projected onto the reaction-plane for semi-central 158 A GeV 
    $^{208}$Pb + $^{208}$Pb collisions (note $y_{cm}=2.9$).
    The vertical errors indicate the statistical errors of the fit
    only. The horizontal bars on the tracking points indicate the
    width of the rapidity bin.
    RQMD model calculations (b=8-10 fm) 
    and VENUS  model calculations (b=8-10 fm) 
    are also shown. The VENUS prediction for $\pi^+$ (not shown) is
    similar to that of RQMD.}
  \label{fig3}
 \end{center}
\end{figure}

Similar to $v_n$, the average projected momenta are reduced  by 
$\langle p^{\prime}_x\rangle = \langle p_x\rangle\cdot\langle 
{\rm cos}(\Phi_0 - \Phi_r)\rangle$.
After correction for the event-plane resolution, the 
$\langle p_x\rangle$ for protons and $\pi^+$ are
plotted as a function of rapidity in Fig.~\ref{fig3}. 
As expected from Fig.~\ref{fig2}, the $\pi^+$ show an anti-flow relative
to the proton flow. Summing over the Plastic Ball acceptance $\langle p_x\rangle$ 
values of $8.2 \pm 0.7, -24.9 \pm 1.9, -53.6 \pm 4.1,$ and $-78.4 \pm
5.8$ MeV/c
are obtained for $\pi^+$, p, d, and t, respectively. The observed scaling
with fragment mass for p, d, and t indicates emission sources with a 
common collective motion.

The main sources of systematic error in the present analysis are:
detector non-uniformities, contamination in the particle
identification, and fit biases in extracting 
$\langle p_x\rangle$~\cite{Theses}.
The mixed event analysis indicates that systematic errors from detector
non-uniformities are less than $2\%$. The effect of contamination has
been estimated by Monte Carlo simulations. The amount of
contamination can be estimated from fitting the background underlying
the peaks in the Plastic Ball particle identification spectra.
For example, the amount of contamination in the proton sample varies
from $6\%$ in peripheral events to $26\%$ in central events. The
effect of contamination has been estimated in simulation by analyzing
events with various amounts and types of contaminated particle 
distributions compared to pure particle distributions, where the 
various distributions are taken to have spectra and flow
characteristics similar to those measured. These studies indicate a
maximum systematic error of $8.5\%$. The extraction of $\langle
p_x\rangle$ is estimated to have an additional $15\%$ uncertainty deduced
from observed variations in the results depending on the fit region or
method used to fit the $d^3N/dp^{\prime}_xdp^{\prime}_ydy$ distribution.
These systematic errors have not been included in Figs.~\ref{fig2} 
or~\ref{fig3}.

In Fig.~\ref{fig3} the measured results are compared to  RQMD~2.3~\cite{Sor90} 
and VENUS~4.12~\cite{Wer93}
predictions for similar impact parameter range. 
The RQMD calculation, in cascade mode, 
overpredicts the observed proton flow by about a factor of three.
On the other hand, at AGS energies cascade mode RQMD
calculations underpredict the observed directed
flow by about a factor of two, but reasonable agreement is obtained when 
mean field effects are included \cite{Bar97}. At SPS energies
mean field effects are expected to be 
smaller, but would increase the observed disagreement.
The VENUS predictions show a similar disagreement in the target
rapidity region. 
The results suggest a significant softness in the nuclear response.
The maximum proton $\langle p_x\rangle$ observed is in better agreement
with predictions of the Quark Gluon String Model (with rescattering)
of Ref.~\cite{Ame91} and with a 3-fluid hydrodynamical model
calculation~\cite{Dum97}. However, 
these predictions have not been filtered
with the experimental acceptance and
both calculations predict that the maximum
$\langle p_x\rangle$ occurs about one unit forward of the target 
rapidity. It is interesting to note that VENUS predicts a complicated proton 
flow behaviour with protons having an anti-flow direction (similar to 
the RQMD pion prediction) near mid-rapidity. However, this prediction  
disagrees with the results of Ref.~\cite{App97}.

In summary, the directed flow of protons and $\pi^+$ has been studied 
in 158 A GeV $^{208}$Pb + $^{208}$Pb collisions. The directed flow
is largest for impact parameter $\approx 8$ fm, which is considerably
more peripheral than observed at lower incident energies. The $\pi^+$
directed flow is in the direction opposite to the protons, similar to
observations at 11 A GeV energy~\cite{Bar97}. The magnitude of the proton
directed flow is much less than
cascade mode RQMD model predictions, which underpredict the proton
flow at AGS energies. It is also much less than
VENUS model predictions. The results indicate a soft
nuclear response compared to these model predictions at SPS energies.

\vspace{0.3cm}
We express our gratitude to the CERN accelerator division for the
excellent performance of the SPS accelerator complex.  We gratefully
acknowledge the effort of all engineers, technicians, and support
staff who have made possible the construction and operation of this
experiment.  

This work is supported by the German BMBF and DFG, the U.S. DOE, the
Swedish NFR and FRN, the Dutch Stichting FOM, the Department of Atomic Energy,
the Department of Science and Technology, and the University Grants
Commission of the Government of India, the Indo-FRG Exchange
Programme, the Stiftung f{\"u}r Deutsch-Polnische Zusammenarbeit,
the Grant Agency of the Czech Republic under contract No. 202/95/0217,
 the PPE division of CERN, the Swiss National Fund,  the INTAS
under contract INTAS-97-0158, the Grant-in-Aid for Scientific Research
(Specially Promoted Research \& International Scientific Research)
of the Ministry of Education, Science and Culture, JSPS Research
Fellowships for Young Scientists
and also by the University of Tsukuba Special Research Projects,
and ORISE. ORNL is managed by Lockheed Martin Energy Research
Corporation under contract DE-AC05-96OR22462 with the U.S. Department
of Energy. MIT is supported by the U.S. Department of Energy under
the cooperative agreement DE-FC02-94ER40818.

\end{document}